\documentclass[prd,nofootinbib,twocolumn,superscriptaddress]{revtex4}

\usepackage{hyperref}

\usepackage{amsfonts,amssymb,amsmath}
\usepackage{epsfig}
\usepackage{mathrsfs}
\usepackage{amssymb}
\usepackage{graphicx}
\graphicspath{ {./images/} }
\usepackage{amsmath}
\usepackage{xcolor}
\usepackage{color}

\hypersetup{hidelinks}

\begin{document}

\title{ Microlensing and event rate of static spherically symmetric wormhole}
\author{Ke Gao}
\email{2021700389@stu.jsu.edu.cn}
\author{Lei-Hua Liu}
\email{liuleihua8899@hotmail.com}

\affiliation{Department of Physics, College of Physics, Mechanical and Electrical Engineering, Jishou University, Jishou 416000, China}

\begin{abstract}
The study focuses on the impact of microlensing in modern cosmology and introduces a new framework for the static spherically symmetrical wormhole in terms of the radial equation of state. Following a standard procedure, the study calculates the lensing equation, magnification, and event rate based on the the radial equation of state. The analysis highlights that the image problem of the light source is complex. Furthermore, the study suggests that larger values for the throat radius of the wormhole and the radial equation of state lead to higher event rates. Additionally, it is proposed that the event rate of a wormhole will be larger compared to that of a black hole, provided their masses and distances from the light source and observer are comparable. This study offers the potential to distinguish between a wormhole and a black hole under similar conditions.

\end{abstract}

\maketitle

\section{Introduction}
\label{introduction}

The first research of WH originated from the inside of Schwarzschild's blackhole \cite{Flamm:1916}. Thereafter, Einstein and Rosen explicitly proposed a vacuum solution that connects two remote regimes \cite{Einstein:1935}. Ref. \cite{Misner:1957} firstly introduced the concept of WH. Then, the most simple WH named the Ellis WH was found by \cite{Ellis:1973} whose ADM mass is zero. To traverse the WH, one kind of traversable WH was introduced by  \cite{Morris:1988cz}. Then, the wormhole was extensively studied by \cite{Damour:2007ap,Kim:2003zb,Bueno:2017hyj,Visser:1989kh,Sushkov:2005kj,Bronnikov:2002rn,Clement:1995ya,Richarte:2017iit,Ayuso:2020vuu,KordZangeneh:2020ixt,Song:2023jdn,Saleem:2023lul,Eid:2023wrd,Godani:2023paj}. The way for realizing the WH, the introduction of exotic matter is mandantory whose energy density is negative, which would violate the Null energy condition (NEC) \cite{Hochberg:1998ii,Hochberg:1998ha,Lobo:2002zf,Lobo:2004rp}.

In the field of astrophysics, the concept of a wormhole (WH) is a hypothetical object predicted by General Relativity. Detecting the existence of wormholes involves observing their lensing effects. Einstein was the first to develop the lensing equation, which introduced the well-known concept of the "Einstein angle" \cite{Einstein:1936llh}. Gravitational lensing has become a standard method to detect various astral objects, including wormholes, dwarf stars, black holes, and more, with the advancement of observation technology. In modern astronomy, lensing includes weak gravitational lensing, strong gravitational lensing, and microlensing. Weak lensing occurs due to the weak potential of the gravitational source, causing only slight distortion of the light passing through the source \cite{Bartelmann:1999yn, Kaiser:1991qi}. On the other hand, strong lensing effects occur when the gravitational source's potential, such as a wormhole or black hole, is strong enough to significantly distort the light \cite{Bozza:2002zj, Virbhadra:1999nm}. Microlensing is an astronomical phenomenon caused by weak gravitational lensing, used to detect objects ranging from planetary mass to stellar mass.
This paper focuses on the microlensing effects. References \cite{Perlick:2004tq, SDSS:2002oin} used the microlensing effect to explore wormholes, leading to extensive research on microlensing effects for wormholes \cite{Gao:2022cds,Liu:2022lfb,Sokoliuk:2022owk,Zatrimaylov:2021ijd,Cheng:2021hoc,Li:2019qyb,Kuniyasu:2018cgv,Raidal:2018eoo,Tsukamoto:2017hva,Sajadi:2016hko,Lukmanova:2016czn,Tsukamoto:2016zdu,Kitamura:2016vad,Kitamura:2012zy,Kitamura:2012wcg,Toki:2011zu,Abe:2010ap,Bogdanov:2008zy,Torres:2001gb,Safonova:2001vz,Torres:1998cu,Torres:1998xd}. Being similar with optics, the gravitational source will bend the light, thus one can observe several images of light source after bending the light. How many images can be formed after bending the light that naturally became a question in the lensing effects, which was discussed in \cite{Kitamura:2016vad,Kuniyasu:2018cgv,Abe:2010ap,Liu:2022lfb}. 

It is important to distinguish between WHs and BHs, especially through observation. In our previous work, we attempted to use magnification to differentiate between them \cite{Gao:2022cds}. However, this method proved to be quite challenging due to the variable distance of the target object between the light source and gravitational source, leading to constantly changing magnification. To explore more possibilities, we plan to employ the equation of state (REoS) to re-formulate the magnification under the Gauss-Bonnet Theorem (GBT). The most crucial factor in this context is the deflection angle. There have been numerous studies related to the deflection angle under GBT \cite{Gibbons:2008rj,Gibbons:2008zi,Werner:2012rc,He:2023hsv,Upadhyay:2023yhk,Javed:2023iih,Javed:2022gtz,Huang:2022soh,He:2022yhp,Gao:2023ltr,Cai:2023ite,Ovgun:2023ego}. We intend to utilize REoS to examine the event rate of white holes, which will be compared with black holes under similar mass and distance conditions between the light source and observer.

This paper is organized as follows: In Sec. \ref{wormhole}, the REoS will be implemented to rewrite the static spherically symmetrical metric. In Sec. \ref{microlensing}, we will use GBT to calculate the deflection angle and lensing equation in light of REoS, meanwhile, the magnification and event rate will be discussed, which are also explicitly related to REoS. In Sec. \ref{conclusion and outlook}, we will give our conclusions and outlook.

\section{Bascis of WH}
\label{wormhole}
In this section, we follow the notation in Refs. \cite{Lobo:2005us, Lobo:2005yv, Garattini:2007ff}. By starting with a spherically symmetical metric, 
\begin{equation}
\label{eq1}
ds^2=-e^{2\Phi}dt^2+\frac{dr^2}{1-b(r)/r}+r^2d\Omega^2,
\end{equation}
which describes a generic static and spherically symmetrical WH metric. By assuming the application of a perfect fluid, its corresponding Einstein equation can be derived as follows, 
\begin{equation}
\label{eq2}
p_r^\prime=\frac{2}{r}\big(p_t-p_r\big)-\big(\rho+p_r\big)\Phi^\prime,
\end{equation}
\begin{equation}
\label{eq3}
b^\prime=8\pi G\rho(r)r^2,
\end{equation}
\begin{equation}
\label{eq4}
\Phi^\prime=\frac{b+8\pi Gp_rr^3}{2r^2\big(1-b(r)/r\big)},
\end{equation}
where the prime denotes a derivative with respect to the radial coordinate $r$, $p_r$ represents the pressure in the radial component, $p_t$ indicates the tangential pressure and $\rho$ is the energy density. REoS is defined as follows,
\begin{equation}
\label{eq5}
p_r=\eta \rho.
\end{equation}
where $\eta$ represents REoS.

The flaring-out condition and asymptotic flatness take the necessary condition:
\begin{equation}
\eta>0 \,\, \text{or} \,\, \eta<-1.
\label{flaring out condition}
\end{equation}
Furthermore, it's important to note that the condition $\eta > 0$ is only possible if $\rho < 0$. Another point to consider is that the Null Energy Condition (NEC) implies $\rho + p_r \geq 0$. By violating the NEC and combining it with the flaring-out condition, we can categorize $\rho$ and $p_r$ into two groups: when $\eta > 0$, it means that $\rho < 0$ and $p_r < 0$, and when $\eta < -1$, it indicates that $\rho > 0$ and $p_r < 0$. This classification is useful for explaining the results presented in Tab. \ref{table:1}.

In light of Eqs. \eqref{eq2}-\eqref{eq5}, one can get
\begin{equation}
\begin{aligned}
b(r) = r_0\bigg(\frac{r_0}{r}\bigg)^\frac{1}{\eta}e^{-(2/\eta)[\Phi(r)-\Phi(r_0)]}\times
\\
\bigg[\frac{2}{\eta} \int_{r_0}^r\big(\frac{r}{r_0}\big)^{(1+\eta)/\eta}\Phi^\prime(r)
e^{(2/\eta)[\Phi(r)-\Phi(r_0)]}dr+1\bigg].
\end{aligned}
\end{equation}
Since we focus on the microlensing effects, it means that the potential of WH is quite weak. For simplicity, one can reasonably assume that $\Phi(r)\approx constant=\Phi(r_0)$. More precisely, the potential varies very slowly from the initial value $\Phi(r_0)$. Therefore, the shape function can be simplified into 
\begin{equation}
b(r)=r_0\bigg(\frac{r_0}{r}\bigg)^{\frac{1}{\eta}}.
\label{simplified impact parameter}
\end{equation}
Substitute this simplified shape function \eqref{simplified impact parameter} into Eq. \eqref{eq1}, then the metric can be transformed towards, 
\begin{equation}
ds^2=-Adt^2+\frac{dr^2}{1-\big(r_0/r\big)^{1+\frac{1}{\eta}}}+r^2d\Omega^2,
\label{simplified metric}
\end{equation}
where $A=e^{2\Phi}$. One can easily observe that metric \eqref{simplified metric} will become an Ellis-Bronnikov WH if the factor $A$ was absorbed into the temporal part and $\eta=1$. Here, we will discuss more about $\eta$. From the example of Ellis-Bronnikov WH corresponding to $\eta=1$, we could see that the various values of $\eta$ will produce the distinctive metric as shown in metric \eqref{simplified metric} that means the different kinds of WHs. 
That is why we approximate $\eta$ as a constant in the calculation since it represents the various kinds of WHs. In the case of $\eta=1$, the simplified metric \eqref{simplified metric} can unify the charged spherically symmetric wormhole \cite{Liu:2022lfb}: $ds^2=-\big(1+\frac{Q^2}{r^2}\big)dt^2+\big(1-\frac{r_0^2}{r^2}+\frac{Q^2}{r^2}\big)^{-1}dr^2+r^2d\Omega^2$ and the Morris-Thorne type wormhole with quantum corrections \cite{Jusufi:2018kmk}: $ds^2=-(1+\frac{\hbar q}{r^2})dt^2+\frac{dr^2}{1-\frac{b_0^2}{r^2}+\frac{\hbar q}{r^2}}+r^2d\Omega^2$ $\it e.t.c$, where these two kinds of WHs can describe many kinds of WH. For instance, Ref. \cite{Liu:2022lfb} shows that the charged spherically symmetric wormhole could desribe 
the  the charged Ellis-Bronnikov WH \cite{Huang:2019lsl} and the generalized Ellis-Bronnikov WH \cite{Bronnikov:1973fh} $\it e.t.c$. Therefore, it is evident that the metric \eqref{simplified metric} can represent various WHs. 
 
 Here, we calculate the $\eta$ in terms of \eqref{eq5} for the case of charged WH whose resulting formula is 
\begin{equation}
\eta=-\frac{r^4 \left(Q^4+Q^2 \left(r^2-r_0^2\right)+r^2 r_0^2\right)}{\left(Q^2+r^2\right)^2 \left(Q^2-r_0^2\right) \left(Q^2+r^2-r_0^2\right)},
\label{eta of charged wh}
\end{equation}
where $r_0$ is the radius of throat and $Q$ represents the charged part. In light of the weak field approximation $r_0\ll r$ (requirement of microlensing), we could obtain $\eta=1$ approximately that is consistent with our analysis. Here, we only consider the case of $\eta=1$, our analysis can also be applied to many other types of WHs corresponding to various values of $\eta$. Based on this analysis, we could consider $\eta$ as a constant in the calculations. 

 Let's delve deeper into $\eta$, which is defined as $\eta=\frac{p_r}{\rho}$. We've already discussed the conditions under which $\eta$ remains constant, namely in the weak field approximation. In the strong field region (where $r_0$ is compatible with $r$), the value of $\eta$ varies, leading to the satisfaction of the NEC. However, the geometry of the Ellis-Bronnikov WH is unique; it consistently violates the NEC throughout the universe since its energy density is negative in light of Einstein equation, whose formulas is 
 \begin{equation}
   \rho=-\frac{1}{8\pi G}\frac{r_0^2}{r^3}. 
   \label{energy denstiy of ellis wh}
 \end{equation}
 And then being armed with $\eta=1$ and $p_r=\rho$, it is evident that the NEC is constantly violated. 
 Although its geometry was discovered first, we only have theoretical motivation to explore it. In the next section, we will implement the REoS to investigate the microlensing effect of the metric \eqref{simplified metric}.

\begin{figure}
    \centering
    \includegraphics[scale=0.7]{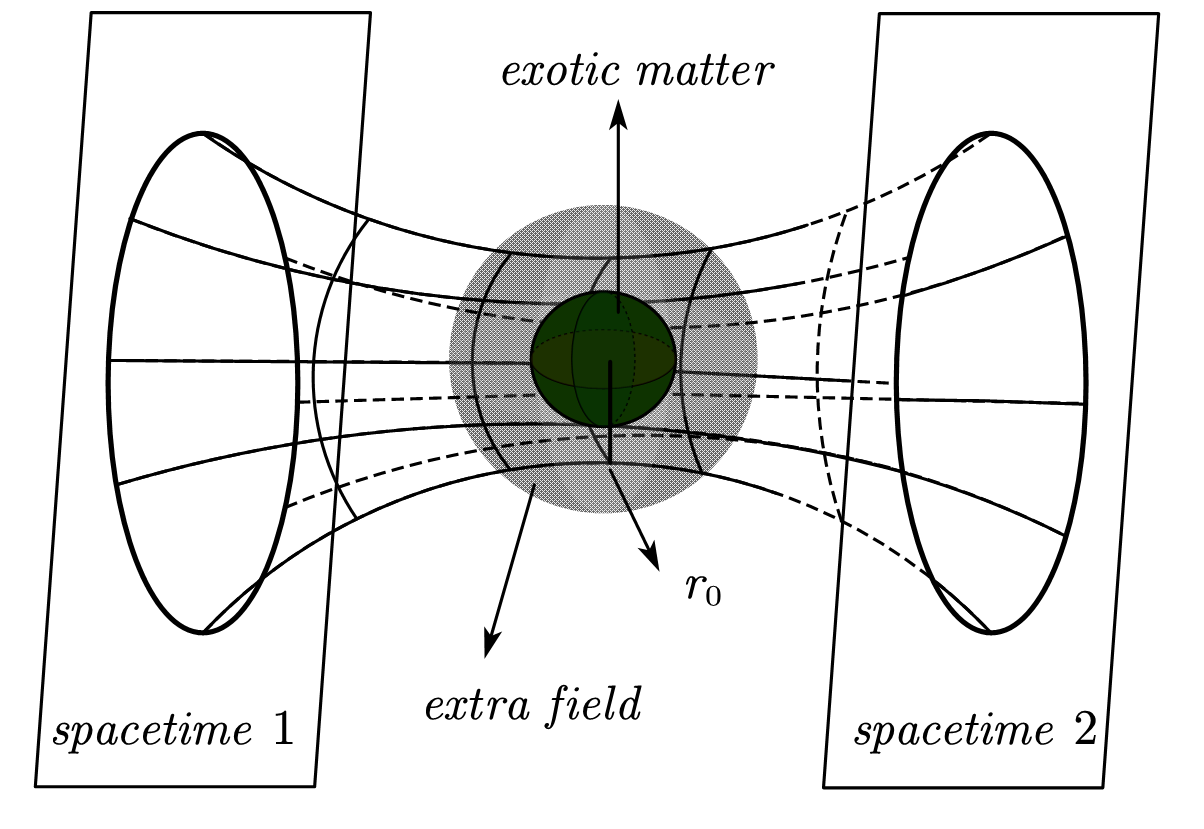}
    \caption{A brief illustration of WH. The WH is connecting two remote regimes of spacetime. In our case, we only consider the lensing effects occurring on one side of WH where is spacetime 1 or spacetime 2. }
    \label{fig:my_label}
\end{figure}

\section{Microlensing}
\label{microlensing}
In this section, we will calculate the magnification and the event rate of metric \eqref{simplified metric}. The calculation will be performed by the GBT.

\subsection{Deflection angle}
\label{deflection angle}
\begin{figure}
	\centering
	\includegraphics[scale=0.8]{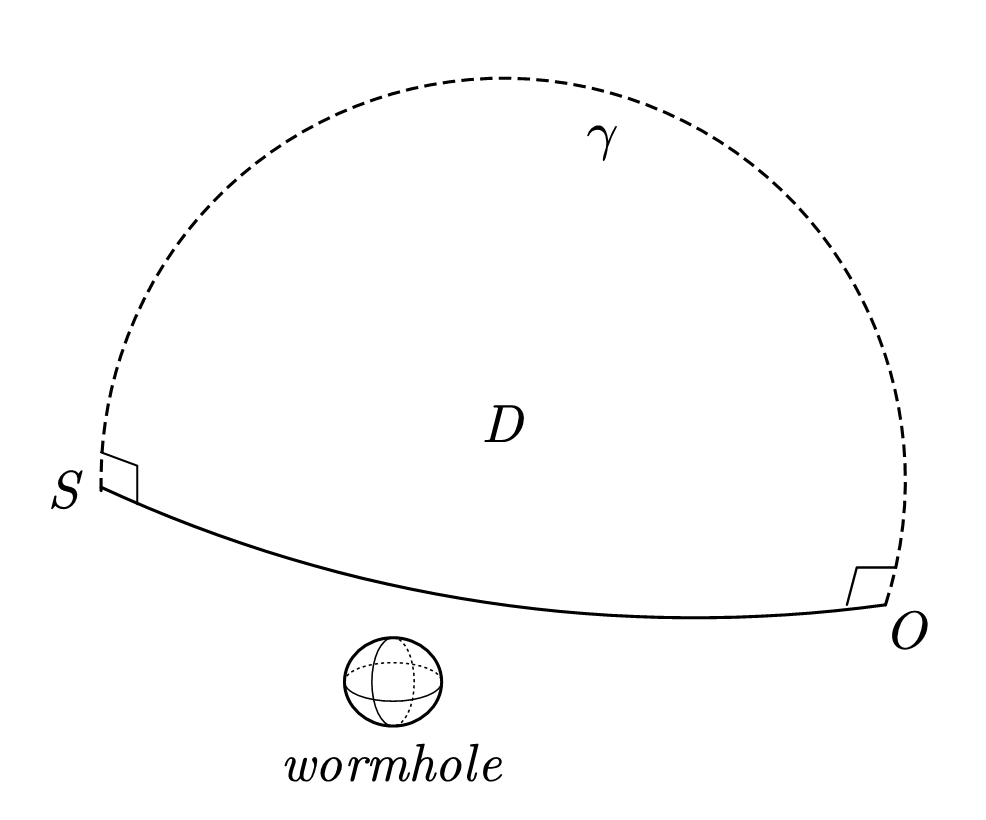}
	\caption{Illustration of the GBT integral domain. $O$ is the observer, and $S$ is the light source. The region $D$ represents the integral domain of the GBT, and integrating over this domain gives us the total deflection angle experienced by the light.}
	\label{fig:1}
\end{figure}

One of the most essential quantify is the deflection angle in lensing effects, which it will be investigated by GBT under the weak field approximation. Before introducing GBT, the concept of optical Gaussian curvature will be given. Considering the light-like case ($ds^2=0$), the metric \eqref{simplified metric} will become 
\begin{equation}
\label{eq 8}
dt^2=\frac{dr^2}{A\left(1-\big(r_0/r\big)^{1+\frac{1}{\eta}}\right)}+\frac{r^2}{A}d\phi^2.
\end{equation}
 Our calculation is performed in the equatorial plane due to the spherical symmetry.
For convinence, we introduce two auxiliary variables:  $du=\frac{dr}{\sqrt{A\big(1-\big(r_0/r\big)^{1+\frac{1}{\eta}}\big)}}$ and $\xi=\frac{r}{\sqrt{A}}$. Gaussian optical curvature can be expressed as, 
\begin{equation}
K=\frac{-1}{\xi(u)}[\frac{dr}{du}\frac{d}{dr}\big(\frac{dr}{du}\big)\frac{d\xi}{dr}+\big(\frac{dr}{du}\big)^2\frac{d^2\xi}{dr^2}],
\end{equation}
combine with metric \eqref{eq 8}, one can get
\begin{equation}
\label{eq10}
K=\frac{-\sqrt{A}r_0\big(\frac{r_0}{r}\big)^\frac{1}{\eta}\big(1+\frac{1}{\eta}\big)}{2r^3\sqrt{1-\big(\frac{r_0}{r}\big)^{1+\frac{1}{\eta}}}}.
\end{equation}
Once obtaining the Gaussian curvature, we could introduce the GBT whose formula is given by 
\begin{equation}
\int\int_D KdS+\int_{\partial D}\kappa dt+\sum\limits_i\alpha_i=2\pi\chi(D),
\label{GBT0}
\end{equation}
where $D$ is the integral domain denoted in Fig. \ref{fig:1} and  $\kappa$ is the curvature of geodesics. Choosing $OS$ as the geodesic line, thus the integral along $OS$ is zero. Additionally, this Euler index $\chi$ is one in domain $D$. Then, GBT \eqref{GBT0} will become 
\begin{equation}
\int\int_{D}KdS+\int_{\gamma_P}\kappa dt+\sum_i\alpha_i=2\pi.
\end{equation}
One can set $\gamma$ to vertically intersect with geodesic line $OS$ at point $O$ and point $S$, Then the sum of external angles is $\pi$ as showing
\begin{equation}
\sum_i\alpha_i=\frac{\pi}{2}(S)+\frac{\pi}{2}(O)=\pi.
\end{equation}
Then, an integral transformation can be done as follows, 
\begin{equation}
\kappa dt=\kappa\frac{dt}{d\phi}d\phi.
\end{equation}
Here $\phi$ is some kinds of angular coordinate where the gravitational source $W$ is the central point as showing in Fig. \ref{fig:2}.  It can be done to set up $\kappa\frac{dt}{d\phi}=1$ on $\gamma$, therefore we have
\begin{equation}
\int\int_{D}KdS+\int_{\phi_O}^{\phi_S}d\phi+\pi=2\pi.
\end{equation}
Geodesic line $OS$ could approximate to be a straight line. Due to the existence of lensing effects, the range of $\phi$ could span from zero (at point $O$) to $\pi+\alpha$ (at point S), where $\alpha$ is the so-called deflection angle,
\begin{equation}
\int\int_{D}KdS+\int_{0}^{\pi+\alpha}d\phi+\pi=\int\int_{D}KdS+\pi+\alpha+\pi=2\pi.
\end{equation}
Then, the deflection angle is obtained as follows, 
\begin{equation}
\alpha=-\int\int_{D}KdS.
\label{deflection angle1}
\end{equation}
Being armed with previous calculations for Gaussian curvature \eqref{eq10}, the deflection angle can be furtherly determined by 
\begin{equation}
\label{eq11}
\alpha=-\int_0^\pi\int_{\frac{b}{\sin\phi}}^\infty K\sqrt{\det  h_{ab}} drd\phi,
\end{equation}
where $b$ is impact parameter and $h_{ab}$ is the optical metric in terms of $u$ and $\phi$ (the induced metric  in coordinates $u$ and $\phi$). Substituting Eq. \eqref{eq10} to \eqref{eq11} and the weak field approximation, one can obtain $\alpha=\frac{\sqrt{\pi}\big(\frac{r_0}{b}\big)^{1+\frac{1}{\eta}}\eta\Gamma[1+\frac{1}{2\eta}]}{2\sqrt{A}\Gamma[\frac{1}{2}\big(3+\frac{1}{\eta}\big)]}$ if $ \frac{1}{\eta}>-2$. As for the special case of $A=1$ and $\eta=1$, our metric becomes the Ellis-Bronnikov WH:
 \begin{equation}
 ds^2=-dt^2+\frac{dr^2}{1-\big(\frac{r_0}{r}\big)^{2}}+r^2d\Omega^2.
 \end{equation}
 The deflection angle ($\eta=1$) is given by
 \begin{equation}
 \alpha=\frac{\pi}{4}\big(\frac{r_0}{b}\big)^2,
 \end{equation}
where it is agreed with Refs. \cite{Gao:2022cds,Nakajima:2012pu,Jusufi:2017gyu} in the first order.

\subsection{Lensing equation}

\begin{figure}
	\centering
	\includegraphics[scale=0.8]{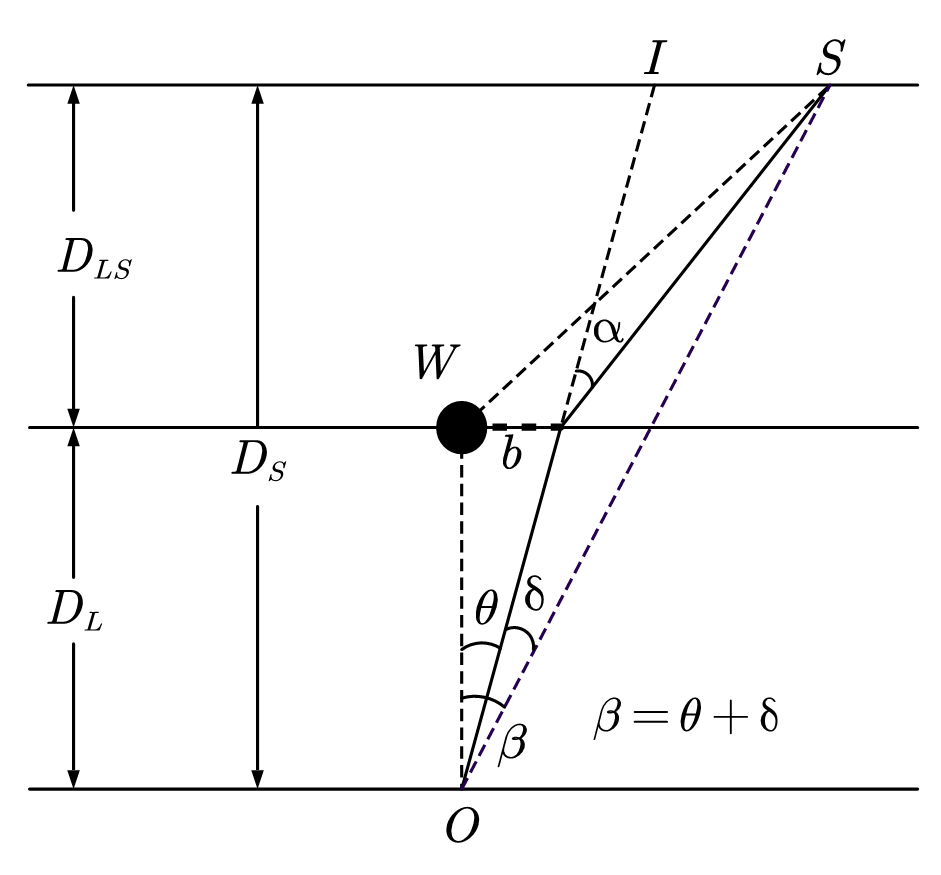}
	\caption{The geometry of lens plane. $I$ is the location of image of light source, $S$ is location of light source, $\alpha$ is the deflection angle, $W$ is the  WH and $b$ is the impact parameter. $\beta$ is the angle between the WH (lensing source) and light source.  $D_{LS}$, $D_L$ and $D_S$ are angular diameter distances.  $\theta$ is the angle between the WH and the image of light source. $\delta$ is the angle between the image of light source and light source. All of these angles are much less than unity.}
	\label{fig:2}
\end{figure}

According to Fig. \ref{fig:2}, the plane geometry could yield the lensing equation
\begin{equation}
\label{eq14}
\beta=\theta-\frac{D_{LS}}{D_S}\alpha.
\end{equation}
Substitute the deflection angle $\alpha$ to Eq. \eqref{eq14}, one can obtain the corresponding lensing equation, 
\begin{equation}
\theta^{2+\frac{1}{\eta}}-\beta\theta^{1+\frac{1}{\eta}}-\frac{D_{LS}}{D_S}\frac{\sqrt{\pi}\big(\frac{r_0}{D_L}\big)^{1+\frac{1}{\eta}}\eta\Gamma[1+\frac{1}{2\eta}]}{2\sqrt{A}\Gamma[\frac{1}{3}\big(3+\frac{1}{\eta}\big)]}=0,
\label{general lens eq}
\end{equation}
where we have used the approximation $b\approx\theta D_L$. According to Eq. \eqref{general lens eq}, one can explicitly obtain the relation between the order of lensing equation (n) and $\eta$, 
\begin{equation}
\label{eq 16}
n=2+\frac{1}{\eta}.
\end{equation}
The above relation is based on the deflection angle without finite distance analysis \cite{Ishihara:2016vdc}, which means that it is an approximated relation. To obtain a more accurate relation with the correction from the finite distance analysis, further work is needed. It is important to note that $n$ should be an integer, limiting the possible values of $\eta$ to correspond to integer values of $n$. It is more beneficial for us to analyze the lensing equation \eqref{general lens eq} in terms of $n$ rather than $\eta$. When $n=2$, there are two images of the light source that correspond to the so-called Einstein ring, which is consistent with $\eta\rightarrow\pm\infty$ (only large values of $\eta$ are considered in this paper). For $\eta=1$, there can be at most three images corresponding to $n=3$, and only one image occurs when $\eta=-1$. We will provide some examples to illustrate that the problems related to images of light sources are complex and highly dependent on $D_L$ and $D_{LS}$. 

 We take $\eta=1$ as an illustration since the charged WH \cite{Kim:2001ri} and WH with quantum corrections \cite{Jusufi:2018kmk}, $\it e.t.c,$ are all this case. Thereafter, lensing equation \eqref{general lens eq} becomes 
\begin{equation}
	\theta^3-\beta\theta^2-M=0,
	\label{lens eq with eta=1}
\end{equation}
where we have set $\frac{D_{LS}}{D_S}=\frac{1}{2}$ for simplicity and $M=\frac{\pi}{16\sqrt{A}\Gamma[4/3]}\frac{r_0^2}{D_L^2}>0$. Lensing equation \eqref{lens eq with eta=1} is the third order equation in terms of $\theta$, in which its overall discriminant $\Delta>0$ in light of appendix I in Ref. \cite{Liu:2022lfb}. Thus, there is only one real solution for Eq. \eqref{lens eq with eta=1} corresponding to one image of the light source. As for the second order of $\theta$, the value of $\eta$ is huge whose value will set to be $10$, then the lensing equation \eqref{general lens eq} will be written by
\begin{equation}
\theta^2-\beta\theta-\frac{5}{2\sqrt{A}}\frac{r_0}{D_L}=0,
\label{second order lensing eq}	
\end{equation}
where we have set $\frac{D_{LS}}{D_S}=\frac{1}{2}$ and the overall discriminant is also larger than zero, there will be two real solutions corresponding to two images of the light source. When $n=4$ and $\sqrt{\frac{\pi}{A}}\frac{1}{8\Gamma[5/3]}\frac{r_0^3}{D_L^3}=0.08$, the lensing equation \eqref{general lens eq} could have four real solutions corresponding to the four images of the light source. For higher order equations of $\theta$ ($n>4$), the situation is more complex, and determining real solutions becomes difficult. Our simple calculation shows that the image problem remains unsolved.

\subsection{Magnification}

Similar to optics, the images of the light source will be magnified or demagnified due to varying the cross-section of light rays, whose definition is determined by 
\begin{equation}
\label{eq 27}
\mu_{\rm total}=\sum_i \bigg|\frac{\beta}{\theta_i}\frac{d\beta}{d\theta_i}\bigg|^{-1},
\end{equation}
where $\theta_i$ is the angle of the $i-th$ image of light  source.
Substituting Eq. \eqref{eq14} into Eq. \eqref{eq 27}, which yields

\begin{figure}
	\center
	\includegraphics[scale=0.72]{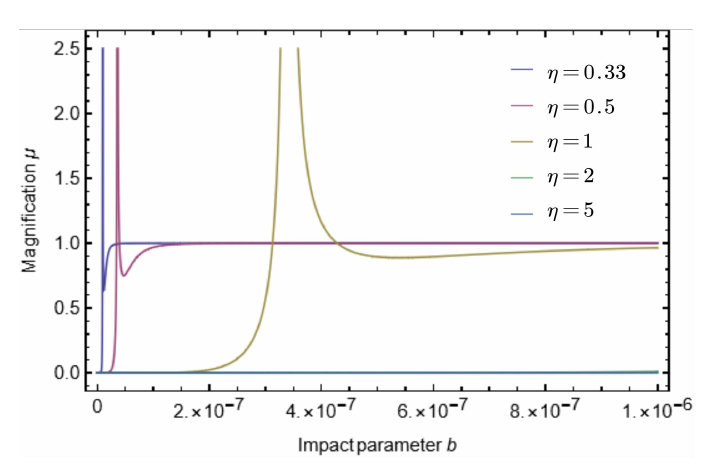}
	\caption{In the case of $\eta>0$, the magnification \eqref{total mag} varies with respect to impact parameter $b$ (Unit: kpc), in which we also consider various values of $\eta$. The parameters are set as follows: $D_S=2D_L=2D_{LS}=20$ kpc, $r_0=1\times 10^{-10}$ kpc, and $A=1$. According to $n=2+\frac{1}{\eta}$, $n=5$ corresponds to $\eta=\frac{1}{3}$, $n=4$ corresponds to $\eta=\frac{1}{2}$ and $n=3$ matches with $\eta=1$. The peak of $\eta=2,5$ is not included since it is beyond the scope.}
	\label{fig: 4}
\end{figure}

\begin{figure}
	\centering
	\includegraphics[scale=0.85]{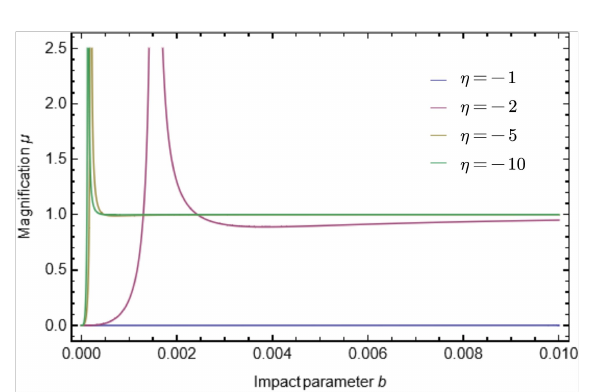}
	\caption{In the case of $\eta<-1$, the magnification \eqref{total mag} varies with respect to impact parameter $b$ (Unit: kpc), in which we also consider various values of $\eta$.  We set the parameters as follows: $D_S=2D_L=2D_{LS}=20$ kpc, $r_0=1\times 10^{-10}$ kpc, and $A=1$. The curves where $\eta=-5$ and $\eta=-10$ have almost overlapped. The case of $\eta=-1$  corresponds to $n=1$ (only one image of light source) whose peak is also not included in this scale.}
	\label{fig: 5}
\end{figure}

\begin{widetext}
\begin{equation}
\mu=\left|\frac{\pi  D_L D_{LS} 2^{-\frac{1}{\eta }-2} r_0 \left(\frac{r_0}{b}\right)^{1/\eta } \left(\sqrt{A} b^2 D_S\Gamma \left(2+\frac{1}{\eta }\right)-D_L D_{LS} 2^{1/\eta } \eta  (\eta +1) r_0 \Gamma \left(1+\frac{1}{2 \eta }\right)^2 \left(\frac{r_0}{b}\right)^{1/\eta }\right)}{A b^4 D_S^2 \Gamma \left(\frac{1}{2} \left(3+\frac{1}{\eta }\right)\right)^2}+1\right|^{-1}.
\label{total mag}
\end{equation}
\end{widetext}
The equation Eq. \eqref{total mag} provides a general formula for calculating the magnification of the static spherically symmetric WH. Due to the flaring-out condition Eq. \eqref{flaring out condition}, the REoS can be divided into two regimes: $(-\infty,-1)$ and $(0,\infty)$. Specific values of $\eta$ have been chosen to investigate the magnification, as shown in Figs. \ref{fig: 4} and \ref{fig: 5}. In these figures, we set $\eta=2$ and $\eta=5$ because we expect that large values of $\eta$ could produce $n=2$ approximately. Another reason is that we could find that the varying trend of magnification of $\mu$ as enhancing $\eta$. Therefore, we only list some large values of $\eta$.

 The trends of magnification in Figs. \ref{fig: 4} and \ref{fig: 5} are similar, each showing only one peak. As the value of $\eta$ increases, the position of the corresponding magnification peak will occur at larger values of $b$. Therefore, the peaks of $\eta=2,5,-1$ are not included in Figs. \ref{fig: 4} and \ref{fig: 5} as they are beyond the scope, and also because there are no magnification effects for these three cases since $\mu=1$. The overall variation trend is as follows: demagnification occurs at a certain scale, followed by an increase in magnification to reach the maximum value (the peak of magnification), and finally it tends towards to one where the size of the image of the light source is the same with the light source itself. The value of $\eta$ can have a significant impact on the position of the magnification peak as it mainly influences the mass of the wormhole, the details of which will be thoroughly investigated in Section \ref{event rate 1}. A simple analysis can help understand this physical picture: the larger the mass of the wormhole, the more significant the distortion of the image of the light source.

\begin{figure}
	\centering
	\includegraphics[scale=0.85]{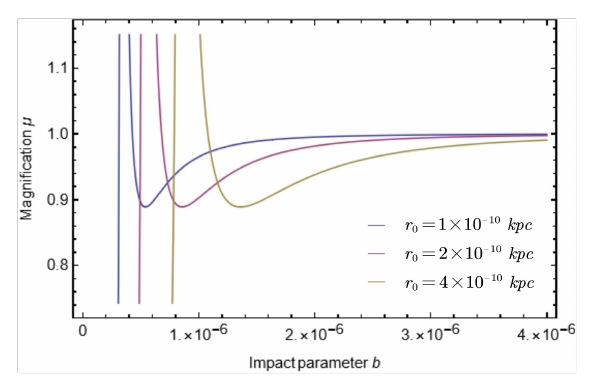}
	\caption{The magnification varies with $b$. We list various values of $r_0$. We set the parameters as follows: $D_S=2D_L=2D_{LS}=20$ kpc, $\eta=1$, $A=1$.}
	\label{fig: 6}
\end{figure}
In order to provide comprehensive information, we have included a graph showing the magnification for various values of $r_0$ in Fig. \ref{fig: 6}. This graph indicates that there is only one peak of magnification. As the values of $r_0$ increase, the peak of magnification will appear at larger values of $b$ since $r_0$ is directly related to the mass of the wormhole (WH). It is important to note that the demagnification of the Kerr black hole is affected by its angular momentum \cite{Johnson:2019ljv,Gralla:2019drh}, and therefore this method can be extended to the study of rotating black holes and wormholes. In summary, it is crucial to recognize that the factor influencing the mass of the wormhole significantly impacts the position of the peak of magnification ($\mu$).

\subsection{Event rate}
\label{event rate 1}
\begin{figure}
	\centering
	\includegraphics[scale=0.8]{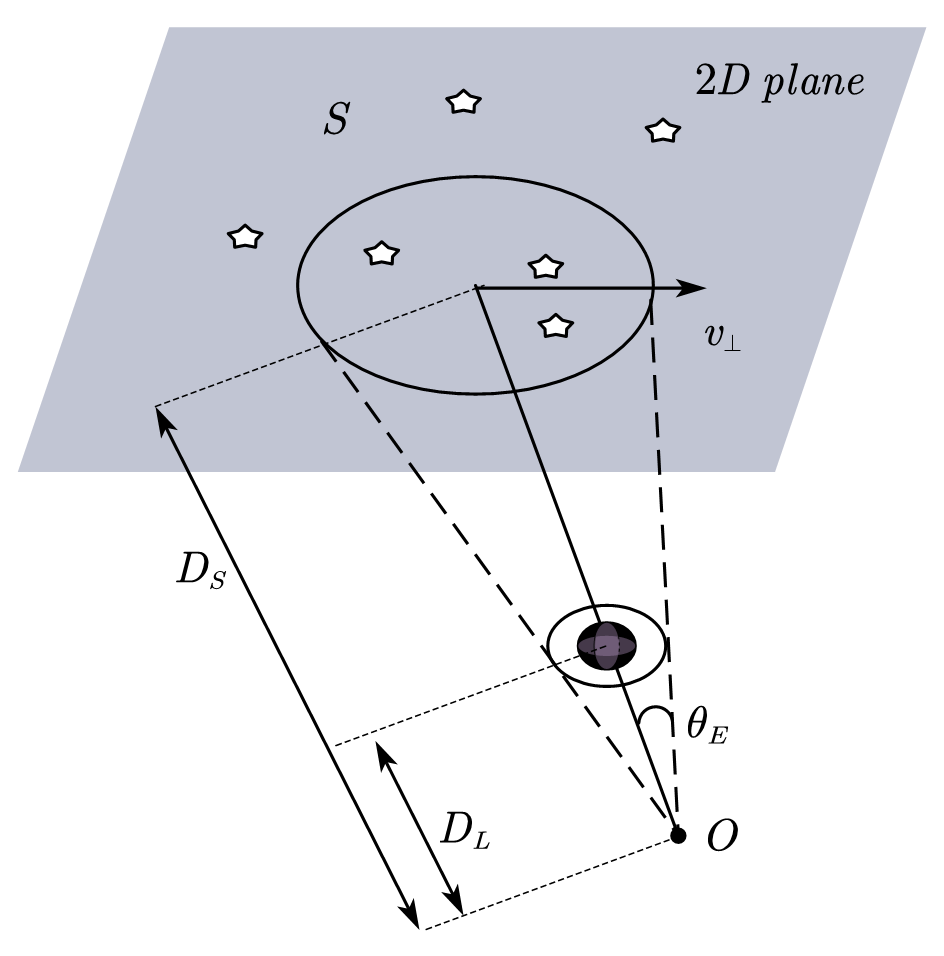}
	\caption{The illustration of the microlensing rate of WH
		moving along this $2D$ plane. This $2D$ plane is the source plane, and we assume that the number of sources within the Einstein ring is $\chi\sigma _{micro}$.}
	\label{fig: 7}
\end{figure}

The microlensing effect is a rare phenomenon in observations. To calculate the probability of this event, we use the concept of optical depth ($\tau$). The optical depth gives us the probability of observing a microlensing event of a source at a specific location $D_S$, which reflects the number of microlensing events per unit of time. If we observe N light sources, we can calculate the microlensing event rate using the formula $\Gamma=\frac{d (N \tau)}{dt}$, where $\tau$ is the optical depth and $t$ is the time. Our goal here is to develop an analytical formula for calculating the event rate based on the simplified metric given by \eqref{simplified metric}. To better understand the event rate, refer to Fig. \ref{fig: 7}.

Our calculation is based on metric \eqref{simplified metric}. First, the effective mass of WH comes via Ref. \cite{Alcubierre:2017pqm} that is defined by  
\begin{equation}
M=\frac{r_0}{2}+\int_{r_0}^r4\pi\rho(r^\prime)r^{\prime 2}dr^\prime,
\end{equation}
where $r$ is the location in the radial direction and the energy density $\rho$ can be found in light of Einstein equation, 
\begin{equation}
\rho=-\frac{Ar_0(\frac{r_0}{r})^{\frac{1}{\eta}}}{r^3\eta}\frac{c^4}{8\pi G}, 
\end{equation}
where $G$ is the Newton constant, $c$ is the speed of light. Following Ref. \cite{Alcubierre:2017pqm}, the effective mass can derived by 
\begin{equation}
M=\frac{A c^4r_0\left(\frac{r_0}{D_L}\right)^{\frac{1}{\eta} }}{2 G}-\frac{A c^4 r_0}{2 G}+\frac{r_0}{2},
\end{equation}
where the integration range is from $r_0$ to $D_L$ as showing in Fig. \ref{fig: 7}. Bofore calculating the event rate, what we need is the Einstein angle whose definition is $\theta_E=\frac{D_{LS}}{D_S}\alpha$ in light of \eqref{eq14}. Then we could plug $\alpha$ into the definition of Einstein angle, one can obtain 
\begin{equation}
\theta_E=\left(\frac{r_0}{D_L}\right)^{\frac{1+\frac{1}{\eta}}{2+\frac{1}{\eta}}}\left(\frac{\sqrt{\pi}\eta\Gamma[1+\frac{1}{2\eta}]}{2\sqrt{A}\Gamma[\frac{1}{2}(3+\frac{1}{\eta})]}\frac{D_{LS}}{D_S}\right)^{\frac{1}{2+\frac{1}{\eta}}},
\end{equation}
where it is explicitly related to Einstein ring where light source, gravitational lensing source and obeserver are aligned. Once obtaining the Einstein angle, one could also define  the cross-section for microlensing as follows, 
\begin{equation}
\sigma_{micro}=\pi\theta^2_E,
\label{solid angle}
\end{equation}
where it is the solid angle producing a detectable microlensing signal. Due to the relative motion between the light source and lens, one can also define the Einstein radius crossing time 
\begin{equation}
\label{te}
t_E=\frac{r_E}{v}=\frac{D_S\theta_E}{v},
\end{equation}
where $r_E$ is the Einstein radius, $v$ is the relative velocity between the light source and the lensing source as shown in Fig. \ref{fig: 7}. For simplification, the lensing source could be fixed and the light source is travelling at speed $v$ whose direction is also showing in Fig. \ref{fig: 7}. The optical depth gives rise to a detectable probability of microlensing event, which defines as follows,
\begin{equation}
\tau=\frac{1}{\Omega}\int_0^{D_S}\sigma_{micro}dN_L,
\end{equation}
 where $N_L$ is the number of microlensing event along the $D_L$ direction. 
As varying with $D_L$, $N_L$ is also changing. Therefore, we could further derive that 
\begin{equation}
dN_L=\Omega D_L^2n(D_L)dD_L,
\end{equation}
where $\Omega$ is the solid angle and $n(D_L)=\frac{\rho}{M}$. Then, the optical depth is 
\begin{equation}
\tau(D_S)=\int_0^{D_S}D_L^2  \frac{\rho(D_L)}{M}\pi \theta_E^2dD_L.
\label{tau}
\end{equation}
The integration interval is $(0, D_S)$, which can be divided into $(D_L,r_0)\cup(r_0,D_{LS})$ according to Fig. \ref{fig:2}. In the interval $(0,r_0)$, the integration of \eqref{tau} is zero since we have neglected the inner structure of WH.  Then, we set parameters as $c=G=A=1$ and $D_S=2D_L=2D_{LS}$. Finally, our integration result is
\begin{equation}
\label{eq. 39}
\left|
\frac{\left(\frac{\sqrt{\pi}\eta\Gamma[1+\frac{1}{2\eta}]}{2\Gamma[\frac{1}{2}(3+\frac{1}{\eta})]}\frac{D_{LS}}{D_S}\right)^{\frac{2}{2+\frac{1}{\eta}}} \left(r_0^{\frac{2(1+\eta)}{1+2\eta}}D_{LS}^{-\frac{2(1+\eta)}{1+2\eta}}-1 \right)   (1+2\eta) } {4\eta(1+\eta)}\right|.
\end{equation}
The obsolute value comes via the positivity of probability. 
We can differentiate the optical depth in light of Fig. \ref{fig:2}, 
\begin{equation}
d\tau=\int_0^{D_S}2n(D_L)r_E^2\frac{dt}{t_E}dD_L, 
\end{equation}
where we have used Eq. \eqref{te}. 
We may observe the microlensing event while we are monitoring a certain number of light sources $N$ (dubbed as constant) within a specific time, its corresponding event rate is defined as
\begin{equation}
\label{eq 41}
\Gamma=\frac{d(N\tau)}{dt}=\frac{2N}{\pi}\int_0^{D_S}n(D_L)\frac{\pi r_E^2}{t_E}dD_L=\frac{2N}{\pi t_E}\tau.
\end{equation}
Substituting the previous calculation results Eq. \eqref{te} and Eq. \eqref{eq. 39} into Eq. \eqref{eq 41}, we obtain

\begin{figure}
    \centering
    \includegraphics[scale=1.0]{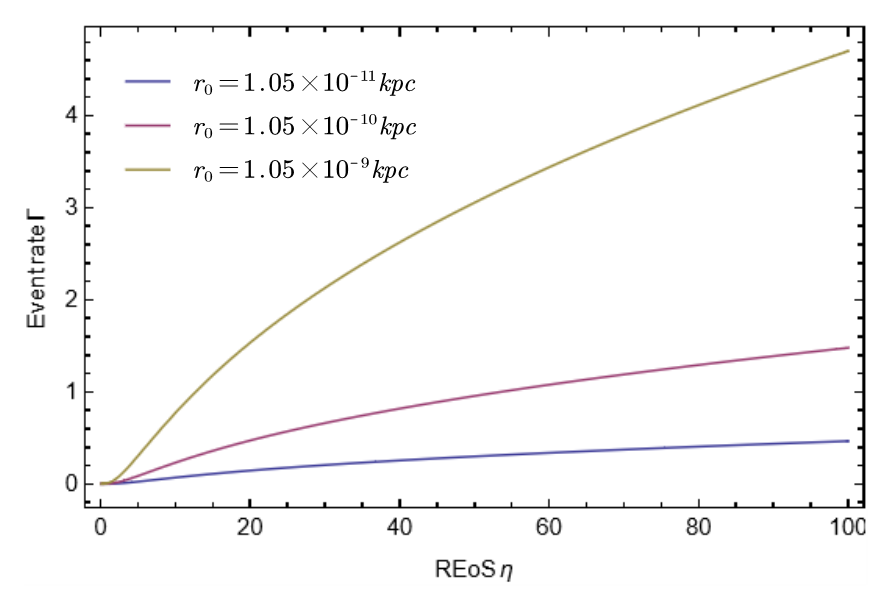}
    \caption{The relationship between the number of microlensing events observed each year and REoS. Each curve in the figure corresponds to a different wormhole throat radius. We set other parameters consistent with Tab \ref{table:1}.}
    \label{fig: event rate}
\end{figure}

\begin{widetext}
\begin{equation}
\label{event rate}
\Gamma=\frac{2\chi \sigma_{micro}}{\pi\frac{D_S}{v}\left(\frac{r_0}{D_L}\right)^{\frac{1+\frac{1}{\eta}}{2+\frac{1}{\eta}}}\left(\frac{\sqrt{\pi}\eta\Gamma[1+\frac{1}{2\eta}]}{2\Gamma[\frac{1}{2}(3+\frac{1}{\eta})]}\frac{D_{LS}}{D_S}\right)^{\frac{1}{2+\frac{1}{\eta}}}}\times
\left|\left(\frac{\sqrt{\pi}\eta\Gamma[1+\frac{1}{2\eta}]}{2\Gamma[\frac{1}{2}(3+\frac{1}{\eta})]}\frac{D_{LS}}{D_S}\right)^{\frac{2}{2+\frac{1}{\eta}}}
\frac{ \left(r_0^{\frac{2(1+\eta)}{1+2\eta}}D_{LS}^{-\frac{2(1+\eta)}{1+2\eta}}-1 \right) (1+2\eta)  } {4\eta(1+\eta)}\right|.
\end{equation}

\begin{table}
\centering
\begin{tabular}{|c |c| c| c| c| c| c| c| c| c|} 
 \hline
 $\eta$ & $\chi$ & $v$ & $r_0$ & $Mass$ & $\theta_E$ & $r_E$ & $t_E$ & $\tau$ & $\Gamma$ \\ 
 - & & $\rm m/s$ & $\rm m$ & $\rm M_\odot$ & $\rm rad$ & $\rm m$ & $\rm year$ & &  $\rm year^{-1}$ \\ 
 \hline
 \hline
  -10 & $1.00\times 10^{14}$ & $3.00\times 10^4$ & $3.24\times 10^{9}$ & $1.38\times 10^7$ & $\mathbb{C}$ & $\mathbb{C}$ & $\mathbb{C}$ & 0.297 & $\mathbb{C}$\\
 -2 & $1.00\times 10^{14}$ & $3.00\times 10^4$ & $3.24\times 10^{9}$ & $3.46\times 10^{11}$ & $\mathbb{C}$ & $\mathbb{C}$ & $\mathbb{C}$ & 0.477 & $\mathbb{C}$\\ 
  -1.1 & $1.00\times 10^{14}$ & $3.00\times 10^4$ & $3.24\times 10^{9}$ & $1.09\times 10^{16}$ & $\mathbb{C}$ & $\mathbb{C}$ & $\mathbb{C}$ & 1.84 &  $\mathbb{C}$\\
0.33 & $1.00\times 10^{14}$ & $3.00\times 10^4$ & $3.24\times 10^{9}$ & $\approx 0$ & $9.67\times 10^{-10}$ & $6.25\times 10^{11}$ & 0.660 & 0.374 & $1.06\times 10^{-4}$ \\
0.5 & $1.00\times 10^{14}$ & $3.00\times 10^4$ & $3.24\times 10^{9}$  & $\approx 0$ & $3.59\times 10^{-9}$ & $2.32\times 10^{12}$ & 2.46 & 0.272  & $2.86\times 10^{-4}$\\
 1 & $1.00\times 10^{14}$ & $3.00\times 10^4$ & $3.24\times 10^{9}$ & 1.09$\times 10^{-5}$ & $3.40\times 10^{-8}$ & $2.20\times 10^{13}$ & 23.2 & 0.201  & $2.00\times 10^{-3}$ \\
  2 &$1.00\times 10^{14}$ & $3.00\times 10^4$ & $3.24\times 10^{8}$ & 0.109 & $5.98\times 10^{-8}$ & $3.87\times 10^{13}$ & 40.9 & 0.187 &   $3.27\times 10^{-3}$\\
 2 &$1.00\times 10^{14}$ & $3.00\times 10^4$ & $3.24\times 10^{9}$ & 3.46 & $2.38\times 10^{-7}$ & $1.54\times 10^{14}$ & 163 & 0.187 &  $1.30\times 10^{-2}$ \\
  2 &$1.00\times 10^{14}$ & $3.00\times 10^4$ & $3.24\times 10^{10}$ & 109 & $9.48 \times 10^{-7}$ & $6.13\times 10^{14}$ & 648 & 0.187 &  $5.19\times 10^{-2}$ \\
 3 & $1.00\times 10^{14}$ & $3.00\times 10^4$ & $3.24\times 10^{9}$ & 236 & $5.92\times 10^{-7}$ & $3.83\times 10^{14}$ & 405 & 0.191 & $3.30\times 10^{-2}$ \\
  10 & $1.00\times 10^{14}$ & $3.00\times 10^4$ & $3.24\times 10^{9}$ & 367 & $6.73\times 10^{-7}$ & $5.84\times 10^{14}$ & 589 & 0.192 & $8.87\times 10^{-2}$ \\
 
 \hline
\end{tabular}
\caption{The event rate $\Gamma$ changes with $\eta$. $\chi$ is the number of sources observed per unit angular area $\pi \theta^2$, $v$ is the relative velocity between the WH and the source plane, $r_0$ is the throat radius of the WH. In terms of mass, we choose solar mass $\rm M_\odot$ as the unit, $\theta_E$ is the Einstein angle, $r_E$ is the Einstein radius, and $\tau$ is the optical depth. We set the parameters $D_S=21~\rm kpc$, and $A=1$. $\mathbb{C}$ means the complex number which is not observed since the $\Gamma$ should be a real number.}
\label{table:1}
\end{table}    

\begin{table}
    \centering
    \begin{tabular}{|c || c| c| c| c| c| c| c|}
    \hline
     $r_0$ ($km$) & $3.24\times 10^1$ & $3.24\times 10^2$ &  $3.24 \times10^3$ &  $3.24 \times10^4$ &  $3.24 \times10^5$ &  $3.24 \times10^6$ &  $3.24 \times10^7$  \\
      \hline
   $\Gamma$ ($year^{-1}$)  &  $1.86\times 10^{-6}$ & $8.62\times 10^{-6}$ & $4.00\times 10^{-5}$ & $1.86\times 10^{-4}$  & $8.62  \times 10^{-4}$ & $4.00 \times 10^{-3}$ & $1.86 \times 10^{-2}$ \\
    \hline
    \end{tabular}
    \caption{The numerical results by Eq \eqref{event rate} : event rate $\Gamma$ of Ellis-Bronnikov WH corresponding to throat radius $r_0$. $D_S=21~\rm kpc$ is assumed. $v=0.0001c$, $\chi=2\times 10^{14}$ and $n(D_L)=\frac{\rho}{M}$ are assumed.}
    \label{tab: 2}
\end{table}

\begin{table}
    \centering
    \begin{tabular}{|c||c| c| c| c| c| c| c|}
    \hline
        $r_0$ ($km$) & 10 & $10^2$ &  $10^3$ &  $10^4$ &  $10^5$ &  $10^6$ &  $10^7$ \\
    \hline
   $\Gamma$ ($year^{-1}$) &  $1.88 \times 10^{-14}$ & $8.73 \times 10^{-14} $ & $4.05  \times 10^{-13}$ & $1.88 \times 10^{-12} $ & $8.73  \times 10^{-12}$ & $4.05 \times 10^{-11}$ & $1.88 \times 10^{-10}$ \\
   \hline
    \end{tabular}
    \caption{The numerical results by \cite{Abe:2010ap}: The various event rates $\Gamma$ of Ellis-Bronnikov WH correspond to different throat radius $r_0$. $D_S=8~\rm kpc$ is assumed. $v=5000~ \rm km/s$ and $n=4.97\times 10^{-9}~\rm pc^{-3}$ are assumed.
}
    \label{tab: 3}
\end{table}

\end{widetext}
where we express $N$ as $\chi\sigma_{micro}$, $\chi\propto D_S^2$ is constant determined by observation. $\chi$ has a significant impact on the event rate that leads to $\Gamma\propto \chi$. Comparing with \cite{Zaris:2019soz,Sollima:2009wh,Noyola:2010ab}, we reasonably set $\chi=1\times 10^{14}$, $D_S=2D_L=2D_{LS}=21~\rm kpc$ and $v=3\times10^{4}~\rm m/s$. Being armed with these parameters, one can determine the mass of WH and Einstein's angle. To illustrate how $\eta$ impacts the event rate, all of our calculations are listed in Tab \ref{table:1}. For similar reasons with Figs. \ref{fig: 4} and \ref{fig: 5}, larger values of $\eta$ will approximately result in $n=2$, where we have set $n=10$ in Tab. \ref{table:1} corresponding to $\eta\approx 2$. When the event rate $\Gamma$ and the Einstein radius $r_E$ are both complex numbers, it indicates that they were unobserved, as the observable should be a real number. As for other values of $\eta$ in Tab. \ref{table:1}, we just show the varying trend of event rate with respect to distinctive $\eta$. It indicates that the event rate and Einstein angle will be enhanced by increasing the value of $\eta$ but not for optical depth. One could see that $\Gamma$ is of order $10^{-2}$ when there are two images per light source. Another point needs to be noticed is that it is meaningless as $\eta<-1$ since the $\Gamma$ is a complex number. Finally, it could be seen that there are at least two images per light source based on $n=2+\frac{1}{\eta}$. For completeness, we also give a plot to show how the radius $r_0$ of WH impacts $\Gamma$ shown in Fig. \ref{fig: event rate}. It explicitly indicates that the event rate will be enhanced by increasing the value of $r_0$. We also show the results of \cite{Abe:2010ap} in Tab. \ref{tab: 3}, where they computed the event rate of clusters in Ellis-Bronnikov WHs with different throat radii. It indicates that the event rate will be enhanced by improving the value of throat radii which supports our numerical results. Their results are different from ours since they consider the cluster of WHs. It leads to the mass becoming sparse which is different from our assumption, where we only consider one WH as the lens source whose energy density is larger. To sum up, the event rate will be larger as increasing the value of $\eta$ and $r_0$.

In addition, we are interested in whether the event rate can be used to differentiate between black holes and wormholes (WHs). In Reference \cite{Kiroglu:2021mej}, they used the CMC Cluster Catalog model to analyze the $n8-rv0.5-rg8-z0.1$ case. In this case, when the mass of the lens is $M=2\times 10^5~\rm M_\odot$ and $t=12 \rm ~Gyr$, the event rate for a single black hole is calculated to be less than $10^{-5}~\rm year^{-1}$. By comparison, we set $\eta=1$ and $r_0=3.24\times 10^{14}~\rm m$, corresponding to $M=1.1\times 10^5~\rm M_{\odot}$. In this scenario, we could detect $5.0\times 10^5$ light sources with an Einstein radius crossing time of $5.0\times 10^5~\rm years$. When the $M$ and $t$ parameters are converted into the $n8-rv0.5-rg8-z0.1$ case, the corresponding event rate is about $10^{-2}\sim 10^{-3}~\rm year^{-1}$. Our estimation reveals that the event rate of a wormhole is two orders of magnitude higher compared to a black hole when the mass and Einstein crossing time are comparable. We would like to emphasize the importance of Tab. \ref{table:1}, where observations can be directly compared with our numerical results by fixing $t$, the mass of the wormhole, and the number of light sources. If the observed values are in line with our predictions, it could indicate that the lensing object is a wormhole. Conversely, if the observed values are smaller than our predictions, it could be indicative of a black hole.

\section{Conclusion and outlook}
\label{conclusion and outlook}
This paper provides a thorough examination of the microlensing effects of the static spherically symmetric WH solution described by Eq. \eqref{simplified metric}. By introducing the REoS parameter $\eta=\frac{p_r}{\rho}$, we reformulate the metric, allowing us to revisit the microlensing effect metric \eqref{simplified metric}, including its magnification and event rate. We utilize the GBT to compute the deflection angle of the metric \eqref{simplified metric} under the weak field approximation. The resulting lensing equation contains an explicit formula $n=2+\frac{1}{\eta}$ without the finite distance analysis, which represents the maximal order of the lensing equation on the equatorial plane. We take $\eta=1$ (corresponding to $n=3$) as an example, where we demonstrate that there is only one real solution of the lensing equation \eqref{general lens eq} since the overall discriminant is greater than zero, indicating that there is only one image of the light source. When $n=2$ ( $\eta=10$ for instance), two real solutions of Eq. \eqref{general lens eq} are observed. Moreover, for $n=4$, the number of images of the light source increases to four, where the values of $\frac{r_0}{D_L}$, $A$, and $D_{LS}/D_S$ have been held constant. However, for higher-order lensing equations, the situation becomes more complex. In some respects, the image problem of the light source remains unresolved due to the complexity of lensing equation \eqref{general lens eq}.

We have reformulated the lensing equation and derived a general formula for calculating magnification in terms of REoS. This formula enables us to analyze how magnification changes with REoS and the WH throat radius ($r_0$). Our analysis reveals that larger values of $\eta$ lead to the position of the magnification's peak occurring at larger values of the impact parameter ($b$), as depicted in Figs. \ref{fig: 4} and \ref{fig: 5}. A similar trend applies to the radius of the WH throat, as illustrated in Fig. \ref{fig: 6}. To provide a comprehensive analysis, we have also conducted analytical calculations of the event rate for a single WH source. Our numerical results are primarily listed in Tab. \ref{table:1} and Fig. \ref{fig: event rate}, indicating that larger values of $\eta$ and $r_0$ result in higher event rates. Our findings align with the investigation of Ref. \cite{Abe:2010ap}, as demonstrated in Tab. \ref{tab: 3}. Specifically, our results in Tab. \ref{table:1} provide a clear comparison with observations by fixing the speed $v$, the mass of WH, and the number of light sources. This serves as a guide for distinguishing WHs from black holes based on event rates.

Our work represents a preliminary investigation of a single gravitational source. The lensing effects can be applicable to more realistic scenarios, such as when the primordial black hole serves as an important component of dark matter. We could extend our method in this direction \cite{Cai:2022kbp}. Another natural extension involves the microlensing of galaxy clusters \cite{Kiroglu:2021mej}, where one can use the approximated metric to simulate their dynamical background. The difference arises from the energy-momentum tensor. Additionally, we could further extend our method to the strong lensing regime, which may encompass the topological effects of spacetime, resulting in different deflection angles and lensing equations. To comprehensively address this issue, the calculation technology needs to be developed.

 \section*{Acknowledgements}
 We appreciate that Hai-Qing Zhang and Bi-Chu Li give lots of suggestions to improve this manuscript. And we are also grateful to Prof. Wentao Luo that he gives professional guidance on some concepts of microlensing. We are very grateful for the critical reading from Dr. Xin-Fei Li. LH and KG are funded by NSFC grant NO. 12165009 and Hunan Natural Science Foundation NO. 2023JJ30487.

\end{document}